\documentclass[twocolumn,aps,prl,superscriptaddress,preprintnumbers,bibnotes]{revtex4-2}

\usepackage{amsmath}
\usepackage{graphicx}
\usepackage{braket}
\usepackage{textcomp}
\usepackage{gensymb}
\usepackage[colorlinks]{hyperref}    % better links - [hidelinks]
\usepackage{xcolor}                                             % for custom link colors
\usepackage{csquotes}                                           % to enquote stuff
\usepackage{ifthen}                                             % for logic in latex

\begin{document}

% custom command to have consistent references for our figures
\newcommand{\figref}[2][]{Fig.~\hyperref[#2]{\ref*{#2}\ifthenelse{\equal{#1}{}}{}{(#1)}}}
\newcommand{\fullfigref}[2][]{Figure~\hyperref[#2]{\ref*{#2}\ifthenelse{\equal{#1}{}}{}{(#1)}}}
\newcommand{\nofigref}[2][]{\hyperref[#2]{\ref*{#2}\ifthenelse{\equal{#1}{}}{}{(#1)}}}
\newcommand{\figsref}[2][]{Figs.~\hyperref[#2]{\ref*{#2}\ifthenelse{\equal{#1}{}}{}{(#1)}}}
% \newcommand{\colorcite}[1]{\textcolor{linkColor}{\cite{#1}}}

% set link colors (remove next lines if unwanted)
\definecolor{linkColor}{HTML}{2e3092}
\hypersetup{
    linkcolor = linkColor,
    citecolor = linkColor,
    urlcolor = linkColor,
}

\title{Fine-Structure Qubit Encoded in Metastable Strontium Trapped in an Optical Lattice}

\newcommand{\AffiliationMPQ}{\affiliation{%
  Max-Planck-Institut f{\"u}r Quantenoptik,
  85748 Garching, Germany}}%
\newcommand{\AffiliationMCQST}{\affiliation{%
  Munich Center for Quantum Science and Technology,
  80799 M{\"u}nchen, Germany}}%
\newcommand{\AffiliationLMU}{\affiliation{%
  Fakult{\"a}t f{\"u}r Physik,
  Ludwig-Maximilians-Universit{\"a}t M{\"u}nchen,
  80799 M{\"u}nchen, Germany}}%

\author{S. Pucher}
\author{V. Kl{\"u}sener}
\author{F. Spriestersbach}
\author{J. Geiger}
\AffiliationMPQ{}
\AffiliationMCQST{}
\author{A. Schindewolf}
\author{I. Bloch}
\author{S. Blatt}
\email{sebastian.blatt@mpq.mpg.de}
\AffiliationMPQ{}
\AffiliationMCQST{}
\AffiliationLMU{}

\date{\today}

\begin{abstract}
We demonstrate coherent control of the fine-structure qubit in neutral strontium atoms.
This qubit is encoded in the metastable $^3\mathrm{P}_2$ and $^3\mathrm{P}_0$ states, coupled by a Raman transition.
Using a magnetic quadrupole transition, we demonstrate coherent state-initialization of this THz qubit.
We show Rabi oscillations with more than 60 coherent cycles and single-qubit rotations on the \textmu{}s scale.
With spin-echo, we demonstrate coherence times of tens of ms.
Our results pave the way for fast quantum information processors and highly tunable quantum simulators with two-electron atoms.
\end{abstract}

\maketitle

Neutral atoms are a promising quantum computing~\cite{saffman2016quantum} and quantum simulation~\cite{gross17} platform due to their long coherence times and highly scalable architecture~\cite{henriet2020quantum, kaufman2021quantum}.
Two-electron atoms in particular have gained increasing attention as their rich level structure offers multiple opportunities to encode high-quality qubits.
Coupling a ground and a metastable state via an optical clock transition has enabled the observation of exceptionally long coherence times~\cite{young20} and direct access to Rydberg states~\cite{madjarov20}.
However, relying on an ultra-narrow optical transition limits operating speed and poses challenges due to an inherent sensitivity to atomic motion and laser phase noise~\cite{chen2022analyzing}.

Faster and more robust qubit rotations can be achieved by coupling two states with a lower energy splitting using a coherent Raman transition~\cite{jenkins2022ytterbium}. Such a coupling scheme has been successfully implemented between nuclear spin states in fermionic isotopes of Yb~\cite{lis2023midcircuit, huie2023repetitive} and Sr~\cite{barnes2022assembly}.
This nuclear-spin qubit led to the experimental demonstrations of high-fidelity gates~\cite{ma2022universal}, erasure conversion~\cite{ma2023high}, and mid-circuit operations~\cite{lis2023midcircuit, huie2023repetitive, norcia2023midcircuit}.
Recently, a complementary encoding of information in electronic degrees of freedom provided by metastable fine-structure states has been proposed~\cite{pagano22, meinert2021quantum}, similar to schemes that have been implemented in ions~\cite{toyoda2009experimental, haze2009measurement}.

In contrast to nuclear spin states, which require a magnetic field to induce a qubit splitting typically in the kilohertz regime~\cite{barnes2022assembly,jenkins2022ytterbium}, the fine-structure states have a natural frequency splitting on the terahertz scale.
Although this splitting makes it more challenging to achieve state-insensitive trapping conditions, it can be advantageous for state preparation and readout, as energy selectivity rather than polarization selectivity can be leveraged~\cite{chen2022analyzing}.
In combination with the existing optical and nuclear qubits, this novel fine-structure qubit can unlock the full potential of the level scheme, leading to new functionalities such as optical qutrits~\cite{trautmann23}, single-photon transition to Rydberg states with fast qubit rotations, and mid-circuit readout operations.

\begin{figure}[b]
\includegraphics{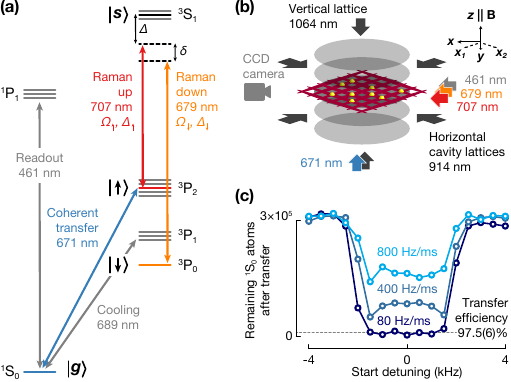}
\caption{Level scheme, schematic of experimental setup, and coherent state initialization. (a) Relevant $^{88}\mathrm{Sr}$ energy levels. The fine-structure qubit is encoded in the metastable states $\ket{{^3\mathrm{P}_2}, m_J = 0}$ and $\ket{^3\mathrm{P}_0, m_J = 0}$, which are coupled via a two-photon Raman transition with a one-photon detuning $\Delta$ from $\ket{{^3\mathrm{S}_1}, m_J = 0}$. The 689-nm and 461-nm light is used for cooling and imaging of the atoms, respectively. (b) Schematic of the experimental setup. Atoms are trapped in a horizontal (vertical) lattice formed at 914\,nm (1064\,nm). The magnetic bias field and the 671-nm state-preparation beam point along the $z$-axis. The Raman beams and the imaging beam propagate along the $x$-axis. (c) Coherent transfer of atoms from $\ket{\mathrm{g}}$ to $\ket{\uparrow}$ using a Landau--Zener sweep with an efficiency of up to 97.5(6)\%. The transfer-laser frequency is swept over 4\,kHz with the indicated ramp speeds.}
\label{fig:setup_landau_zener}
\end{figure}

Here, we experimentally demonstrate core capabilities of a fine-structure qubit using Sr atoms trapped in an optical lattice. As shown in \figref[a]{fig:setup_landau_zener}, our qubit is encoded in the metastable triplet states $\ket{\uparrow} = \ket{5\mathrm{s}5\mathrm{p}\,{^3\mathrm{P}_2}, m_J = 0}$ and $\ket{\downarrow} = \ket{5\mathrm{s}5\mathrm{p}\,{^3\mathrm{P}_0}, m_J = 0}$, which are separated by about $17$\,THz in frequency. These states are coupled via a two-photon Raman transition through the triplet state  $\ket{\mathrm{s}} = \ket{5\mathrm{s}6\mathrm{s}\,{^3\mathrm{S}_1}, m_J = 0}$. We demonstrate fast two-photon Rabi oscillations with frequencies up to $2 \pi \times 400$\,kHz and study their decoherence mechanisms.
We show proof-of-principle read-out methods with about 96\% detection efficiency that can be used for mid-circuit read-out.
Finally, we investigate the coherence of the fine-structure qubit with Ramsey and spin-echo measurements.

The experimental setup is illustrated in \figref[b]{fig:setup_landau_zener}.
We load about $10^{5}$ $^{88}\mathrm{Sr}$ atoms into a 3D optical lattice~\cite{klusener24}.
In the horizontal direction, the atoms are trapped using laser light with a wavelength of 914\,nm, which forms an optical lattice inside an enhancement cavity~\cite{park22}. A retro-reflected laser beam at 1064\,nm generates the vertical lattice. After loading the atoms into the lattice, we perform resolved sideband cooling on the $^{1}\mathrm{S}_0$--$^{3}\mathrm{P}_1$ transition at a horizontal (vertical) lattice depth of $150\,E_\mathrm{rec}$ ($270\,E_\mathrm{rec}$), where $E_\mathrm{rec} = h^2/(2m\lambda_\mathrm{l}^2)$ is the lattice photon recoil energy for an atom with mass $m$ at the corresponding lattice wavelength $\lambda_\mathrm{l}$, and $h$ denotes the Planck constant. The trap depth corresponds to a horizontal (vertical) trap frequency of $65\,\mathrm{kHz}$ ($68\,\mathrm{kHz}$).
We typically achieve temperatures of about $2.5$\,\textmu{}K~\cite{klusener24}.
To initialize the qubit in $\ket{\uparrow}$, we coherently excite the atoms from the ground state $\ket{\mathrm{g}} = \ket{5\mathrm{s}^2\,{^1\mathrm{S}_0}}$ to $\ket{\uparrow}$ using the magnetic field and lattice parameters from Ref.~\cite{klusener24}.
To achieve a robust state preparation, we perform a Landau--Zener sweep with a typical transfer efficiency of 97.5(6)\%, as shown in \figref[c]{fig:setup_landau_zener}~\cite{supplemental}.
The same sweep is also used for state-selective readout of $\ket{\uparrow}$.

After state preparation, we set the magnetic field to 20\,G, oriented along the $z$-direction, corresponding to a Zeeman splitting of 42\,MHz in the ${^3\mathrm{P}_2}$ manifold, which helps to isolate the $m_J=0$ state. We refer to the Raman laser beams driving the $\ket{\uparrow}$--$\ket{\mathrm{s}}$ and the $\ket{\mathrm{s}}$--$\ket{\downarrow}$ transition as the up and down lasers, respectively.
They are both $\pi$ polarized and co-propagate in the $x$-direction to minimize momentum transfer, with a Lamb-Dicke parameter of $0.01$.
We stabilize the laser frequencies to a shared optical reference cavity to ensure phase stability between the lasers.
To suppress spontaneous decay from $\ket{\mathrm{s}}$, we can detune the Raman lasers from the atomic transition frequencies by the one-photon detuning~$\Delta = \Delta_{\uparrow} \approx \Delta_{\downarrow}$, where $\Delta_{\uparrow}$ and $\Delta_{\downarrow}$ are the detunings of the up laser and the down laser, respectively. The two-photon detuning $\delta = \Delta_{\uparrow} - \Delta_{\downarrow}$ is typically set to zero.

To achieve long coherence times within the $\Lambda$ system, it is necessary to mitigate differential light shifts of the qubit states.
The non-spherical $\ket{\uparrow}$ state features a tensor polarizability which allows tuning its light shift relative to the light shift of $\ket{\downarrow}$ by tilting the linear polarization of the trapping light field with respect to the magnetic quantization axis~\cite{trautmann23}.
For the horizontal 914-nm lattice, we find the so-called \enquote{magic} trapping condition where the differential polarizability between $\ket{\uparrow}$ and $\ket{\downarrow}$ vanishes, close to the theoretically predicted angle of about $79\degree$. For the vertical 1064-nm lattice no such magic angle exists and a residual differential polarizability on the percent level remains~\cite{supplemental}.
To minimize the differential light shift, we reduce the vertical lattice depth to about $27\,E_\mathrm{{rec}}$ and choose its polarization to be orthogonal to the magnetic field.
The gravitational tilt of the vertical lattice helps to suppress tunneling and collisions, which are negligible on our experimental time scales~\cite{lemonde2005optical, traverso09}.

\begin{figure}
\includegraphics{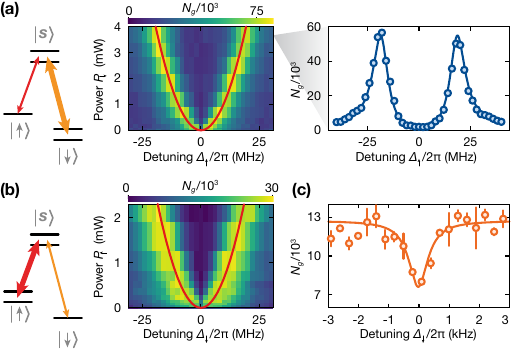}
\caption{
Characterization of the $\Lambda$ system.
(a)~Autler--Townes splitting probed on the $\ket{\uparrow}$--$\ket{\mathrm{s}}$ transition in the presence of a strong resonant down-laser field, as illustrated by the level scheme.
The color scale indicates the number of atoms that decayed into $\ket{\mathrm{g}}$ through $^3\mathrm{P}_1$ after excitation to $\ket{\mathrm{s}}$.
The blue data points show an example of a spectrum for fixed down-laser power $P_{\downarrow} = 3.6$\,mW, which is fitted with an electromagnetically induced transparency model~\cite{fleischhauer2005eit} (blue line) to extract the level splitting of $\ket{\mathrm{s}}$. The red line represents a fit of the level splittings used to calibrate the one-photon Rabi frequency of the down transition.
(b)~The analog calibration of the one-photon Rabi frequency of the up transition.
(c)~Excitation spectrum at low coupling strength with resonant up-laser field. A Lorentzian fit (solid line) to the narrow dip in the data with a full width at half maximum of $2\pi\times 0.71(19)$\,kHz demonstrates a large degree of coherence in our system.
}
\label{fig:autler_townes_cpt}
\end{figure}

\begin{figure*}
\includegraphics{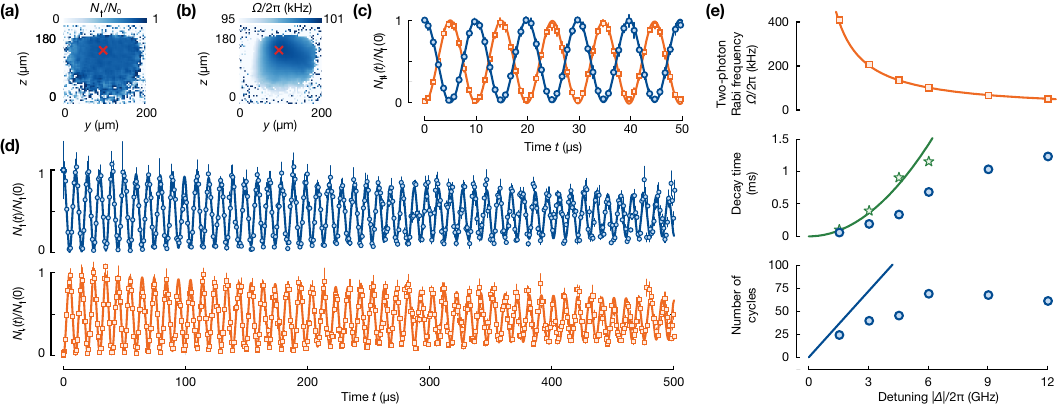}
\caption{
  Fine-structure qubit Rabi oscillations. (a) Measurement of the spatial distribution of the number of atoms in $\ket{\uparrow}$. (b) Spatial distribution of two-photon Rabi frequency $\Omega$. The data for the Rabi oscillations are taken at the red cross. (c) Rabi oscillations with Raman lasers about $6$\,GHz red-detuned from $\ket{\mathrm{s}}$. Blue circles (orange squares) show the number of atoms $N_\uparrow(t)$ and $N_\downarrow(t)$ in $\ket{\uparrow}$ and $\ket{\downarrow}$, respectively, both normalized with interleaved measurements of $N_\uparrow(0)$ and averaged over 10~experimental runs. The lines are fits with a sinusoidal oscillation. (d) A similar measurement as shown in (c), with data from individual runs of the experiment.
The lines represent a damped sinusoidal-oscillation model illustrating that $\Omega = 2\pi \times 100.92(4)$\,kHz and the exponential decay time $\tau = 0.68(1)$\,ms, both found with our analysis method~\cite{supplemental}, provide a good description of the data.
The error bars correspond to the standard deviation of the data in a $3 \times 3$ pixel region. (e) Rabi oscillations at various one-photon detunings $\lvert\Delta\rvert/2\pi$. Upper panel: the Rabi frequency (orange squares) shows a $1/\Delta$ dependence (orange line). Middle panel: The $1/e$ decay time $\tau$ of the envelope of the oscillations (blue dots) increases with $\Delta$. The green stars and the solid line show the limit of the decay time given by the one-photon scattering rate. Lower panel: the number of cycles $\Omega\tau/(2\pi)$ (dots) increases with $\Delta$ and saturates at about 69~cycles (see text). The blue line shows the scattering-limited number of cycles inferred from the fits in the upper and middle panels.
}
\label{fig:rabi}
\end{figure*}

To characterize the $\Lambda$ system $\ket{\uparrow}$--$\ket{\mathrm{s}}$--$\ket{\downarrow}$ containing the qubit subspace, we perform Autler--Townes spectroscopy.
With this method, we calibrate the Rabi frequencies and demonstrate coherent population trapping to reveal the presence of coherence in the system.
First, we resonantly couple $\ket{\downarrow}$ and $\ket{\mathrm{s}}$ by applying the down laser with variable power $P_\downarrow$.
The resulting splitting is given by the Rabi frequency $\Omega_{\downarrow}\propto\sqrt{P_\downarrow}$ of the down-laser field~\cite{autler1955stark}.
We probe this splitting by scanning the detuning $\Delta_{\uparrow}$ at $P_{\uparrow} = 30$\,{\textmu}W, as illustrated in \figref[a]{fig:autler_townes_cpt}.
Readout is performed via detection of atoms that are excited from $\ket{\uparrow}$ to $\ket{\mathrm{s}}$ and subsequently decay through $^3\mathrm{P}_1$ into $\ket{\mathrm{g}}$.
A fit of the data results in $\Omega_{\downarrow}/\sqrt{P_{\downarrow}} = 2\pi\times 19.3(1)\,\mathrm{MHz}/\sqrt{\mathrm{mW}}$.

Next, we prepare the atoms in $\ket{\downarrow}$ and repeat the measurement with exchanged roles of the laser fields. We use $P_\downarrow \approx 220$\,nW and find $\Omega_{\uparrow}/\sqrt{P_{\uparrow}} = 2\pi\times 24.3(2)\,\mathrm{MHz}/\sqrt{\mathrm{mW}}$, see \figref[b]{fig:autler_townes_cpt}. Finally, to observe coherent population trapping~\cite{agapev93, arimondo96, finkelstein2023practical}, we reduce the power in both laser fields significantly to about $10$\,nW. At the two-photon resonance we observe a narrow dip in the excitation spectrum, as shown in \figref[c]{fig:autler_townes_cpt}. A Lorentzian fit yields a full width at half maximum of $2 \pi \times 0.71(19)$\,kHz, four orders of magnitude narrower than the excited state's inverse lifetime $2 \pi \times11$\,MHz~\cite{heinz20}.
This feature demonstrates the presence of coherence in the $\Lambda$ system.

We now demonstrate coherent control of the fine-structure qubit by performing two-photon Rabi oscillations.
To this end, we prepare the atoms in $\ket{\uparrow}$, set the one-photon detuning to $\Delta \approx -2 \pi \times 6$\,GHz, and tune the lasers to the two-photon resonance $\delta = 0$.
Both laser fields have a Rabi frequency of $\Omega_{\uparrow} = \Omega_{\downarrow} \approx 2 \pi \times 36$\,MHz to minimize differential light shifts.
After driving two-photon Rabi oscillations with frequency $\Omega$ for a variable time $t$, we perform a state-selective readout of atoms in $\ket{\uparrow}$.
For this purpose, we perform another Landau--Zener sweep to transfer the population from $\ket{\uparrow}$ to $\ket{\mathrm{g}}$.
We then detect the number of atoms in $\ket{\mathrm{g}}$ in a spatially resolved manner with absorption imaging on the $^{1}\mathrm{S}_0$--$^{1}\mathrm{P}_1$ transition, as shown in \figref[a]{fig:rabi}.
We achieve a detection fidelity of atoms in $\ket{\uparrow}$ of $(95.7 \pm 2.8)\%$, currently limited by the efficiency of the Landau--Zener sweep~\cite{supplemental}.
We repump any remaining atoms in $\ket{\uparrow}$ and remove them from the trap.
This procedure leads to less than 1\% contamination of the $\ket{\downarrow}$ population with atoms from $\ket{\uparrow}$.
Then, we detect the number of atoms in $\ket{\downarrow}$ by repumping them to $\ket{\mathrm{g}}$ via $\ket{\mathrm{s}}$ using both Raman lasers. We take another absorption image and estimate a detection fidelity for atoms in $\ket{\downarrow}$ of $(95.9 \pm 3.3)\%$~\cite{supplemental}.

When we analyze the Rabi frequencies spatially resolved, we find a variation of the Rabi frequencies due to the finite beam size of the Raman lasers in the $yz$-plane, see \figref[b]{fig:rabi}.
To minimize the influence on the dephasing of the Rabi oscillations, we analyze the data at one spatial location~\cite{supplemental}.
To correct for long-term drifts of the atom number, we interleave reference measurements of the atom number in the initial state $\ket{\uparrow}$, which we use to normalize the population in $\ket{\uparrow}$ and $\ket{\downarrow}$~\cite{supplemental}.
We observe high-contrast Rabi oscillations in $\ket{\uparrow}$, as shown in \figref[c]{fig:rabi}.
For a $\pi$ pulse with duration of about 5\,{\textmu}s, we find an excitation fraction of 98(1)\%~\cite{supplemental}, without including the state-detection efficiencies above.

Next, we investigate the long-term dynamics of the Rabi oscillations, as shown in \figref[d]{fig:rabi}.
To this end, we analyze the data to extract both carrier frequency and envelope of the oscillations in $\ket{\uparrow}$ separately.
We determine the carrier frequency corresponding to $\Omega$ from a Lorentzian fit to the Fourier transform of the data~\cite{supplemental}.
Then, we apply a band-pass filter around $\Omega$ and detect the envelope of the Rabi oscillations.
We fit the envelope of the filtered data with an exponential function to find the $1/e$ decay time $\tau$ of the envelope~\cite{supplemental}.
An exponentially damped sinusoidal-oscillation model based on $\Omega$ and $\tau$ obtained from this analysis method yields good agreement with our data of both qubit states, as shown in \figref[d]{fig:rabi}~\cite{supplemental}.

Now, we study the influence of the one-photon detuning $\Delta$ on the Rabi oscillations, see \figref[e]{fig:rabi}.
We measure Rabi frequencies up to $\Omega = 2\pi \times 409(1)$\,kHz and find an expected scaling of the Rabi frequency as $\Omega\propto1/\Delta$. The resulting decay times $\tau$ at small $\Delta$ are limited predominantly by one-photon scattering, which we characterize with independent measurements of the scattering rate, see middle panel of \figref[e]{fig:rabi}~\cite{supplemental}.
We find that the number of cycles increases with $\lvert \Delta \rvert$ and saturates at about 69~cycles, as shown in the lower panel of \figref[e]{fig:rabi}~\cite{supplemental}.
The scaling of $\tau$ with $\Omega$ and the saturation to 69 cycles is consistent with a residual inhomogeneity of $\Omega$ on the order of 0.4\%, presumably caused by laser intensity noise and residual spatial inhomogeneity.
We note that effects due to laser intensity noise could be strongly suppressed with standard composite pulse sequences~\cite{bluvstein2022quantum}.

\begin{figure}
\includegraphics{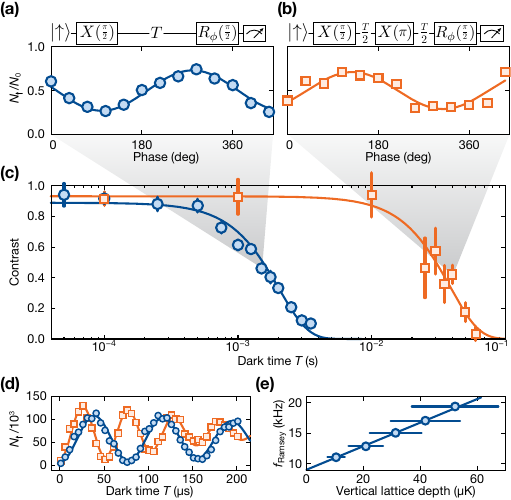}
\caption{Coherence measurements of the Sr fine-structure qubit.
(a) Ramsey measurement. Two $\pi/2$ pulses are applied with a dark time $T$ in between. The phase of the second $\pi/2$ pulse is scanned. The resulting oscillations of the $\ket{\uparrow}$ population (dots) are fitted with a sinusoidal function (line). (b) Spin-echo measurement. A $\pi$ pulse at $T/2$ lets the spins rephase at $T$.
(c) Ramsey (blue dots) and spin-echo (orange squares) measurements for various $T$. We extract the contrast from the fits in (a) and (b) and fit a Gaussian decay with $1/e$ decay times of $T_{2}^* = 2.03(7)$\,ms (blue line) and $T_2' = 38(3)$\,ms (orange line). (d) Ramsey fringes as a function of $T$ to extract the differential lattice light shift for vertical lattice depths of 21(6)\,\textmu{}K (blue dots) and 52(15)\,\textmu{}K (orange squares), as experienced by $\ket{\uparrow}$. The Raman lasers are tuned to a two-photon detuning of about $2\pi \times 10$\,kHz with respect to the free-space resonance. The frequency $f_{\mathrm{Ramsey}}$ of the $\ket{\uparrow}$ population oscillations over the dark time $T$ is determined with a damped sinusoidal fit model (lines). (e) $f_{\mathrm{Ramsey}}$~for different vertical lattice depths of $\ket{\uparrow}$. A linear fit yields a differential light shift of 192(82)\,Hz/\textmu{}K between $\ket{\uparrow}$ and $\ket{\downarrow}$ at the lattice-light wavelength of $1064$\,nm.
}
\label{fig:ramsey_spin_echo}
\end{figure}

Finally, to investigate the qubit coherence we perform Ramsey and spin-echo experiments, see \figsref[a]{fig:ramsey_spin_echo} and~\nofigref[b]{fig:ramsey_spin_echo}.
We apply a $\pi / 2$ pulse to prepare a coherent superposition state $\frac{1}{\sqrt{2}}(\ket{\uparrow} + \ket{\downarrow})$. After a dark time $T$, we map the coherence onto population oscillations by varying the phase of a second $\pi / 2$ pulse.
In \figref[c]{fig:ramsey_spin_echo}, we show the decay of the Ramsey contrast with increasing $T$. From a Gaussian fit to the data, we extract a $1/e$ dephasing time of $T_{2}^* = 2.03(7)$\,ms, limited by residual differential light shifts of the non-magic vertical lattice.

To quantify this light shift, we measure the change in the $\ket{\uparrow}$--$\ket{\downarrow}$ transition frequency as a function of the vertical lattice depth using Ramsey spectroscopy.
Contrary to the Ramsey measurements above, we do not scan the phase of the second $\pi/2$-pulse, but the dark time $T$ between the pulses. The Raman laser frequencies are set to a two-photon detuning of $\delta \approx 2\pi \times 10\,\mathrm{kHz}$ with respect to the free-space resonance of the qubit. The differential light shift that the atoms experience in the lattice causes an additional change in $\delta$.
The frequency of the resulting Ramsey oscillations $f_\mathrm{Ramsey}$ is equal to $\delta/(2\pi)$ and is extracted from a fit with a sinusoidal function, see \figref[d]{fig:ramsey_spin_echo}. \fullfigref[e]{fig:ramsey_spin_echo} shows the dependence of the Ramsey frequency on the vertical lattice depth.
A linear fit to the data extracts a differential light shift of 192(82)\,Hz/\textmu{}K between $\ket{\uparrow}$ and $\ket{\downarrow}$ at a vertical lattice wavelength of $1064\,\mathrm{nm}$, corresponding to a differential polarizability of about 1\%.

To reduce dephasing caused by this differential light shift and other slow fluctuations present in the system, we carry out spin-echo measurements.
We add a $\pi$ pulse after $T/2$ to the Ramsey sequence, which lets the spins rephase at $T$.
With this method, we extend the contrast decay time to $T_2' = 38(3)$\,ms, as shown in \figref[c]{fig:ramsey_spin_echo}.
We project an additional order-of-magnitude increase in coherence time in a 3D magic lattice, supported by our recent results for the $\ket{\mathrm{g}}$ and $\ket{\uparrow}$ states~\cite{klusener24}.

In summary, we demonstrated a new fine-structure qubit encoded in metastable Sr operating at a qubit splitting of 17\,THz.
We presented coherence times of tens of milliseconds, orders of magnitude longer than the single qubit gate times on the microsecond scale.
Extending the coherence times further by an order of magnitude should be possible by modifying the currently limiting wavelength of the vertical lattice~\cite{klusener24,trautmann23}.
Particularly promising is a trapping wavelength of 813\,nm, at which, based on our experimental results, we predict a triple magic condition for both qubit states and the ground state~\cite{trautmann23,supplemental}.
In such a configuration, the $^1\mathrm{S}_0$--$^3\mathrm{P}_0$ clock transition and the $^1\mathrm{S}_0$--$^3\mathrm{P}_2$ transition, used here for coherent transfer, could serve as additional tools for state-selective shelving operations and mid-circuit readout. Moreover, the manipulation of the all-optical qutrit $^1\mathrm{S}_0$--$^3\mathrm{P}_0$--$^3\mathrm{P}_2$ will be possible. Alternatively, operation at a triple-magic trap wavelength for the fine-structure qubit states and a particular Rydberg state might prove beneficial for reaching high two-qubit gate fidelities~\cite{meinert2021quantum,pagano22}. Finally, fast state-selective readout of the fine-structure qubit states without the requirement of a slow shelving pulse can be implemented using the 5s5d\,$^3$D states~\cite{daley08, stellmer2014reservoir}.

\begin{acknowledgments}
  In a study performed in parallel to ours, similar results have been achieved with atoms trapped in optical tweezers~\cite{unnikrishnan24}. We thank M.\,Ammenwerth, F.\,Gyger, F.\,Meinert, T.\,Pfau, R.\,Tao, J.\,Trautmann and J.\,Zeiher for stimulating discussions, S.\,Snigirev from planqc for support with experimental control hardware and software, and D.\,Filin and M.\,S.\,Safronova for providing the polarizability data. We acknowledge funding by the Munich Quantum Valley initiative as part of the High-Tech Agenda Plus of the Bavarian State Government, by the BMBF through the program ``Quantum technologies -- from basic research to market'' (Grant No. 13N16357), and funding under the Horizon Europe program HORIZON-CL4-2022-QUANTUM-02-SGA via the project 101113690 (PASQuanS2.1). V.\,K. thanks the Hector Fellow Academy for support.
\end{acknowledgments}

% \bibliography{finestructure}

\end{document}

% custom command to have consistent references for our figures
% \newcommand{\figref}[1]{\hyperref[#1]{Fig.~\ref*{#1}}}
% \newcommand{\figref}[1]{Fig.~\hyperref[#1]{\ref*{#1}}}
\newcommand{\figref}[2][]{Fig.~\hyperref[#2]{\ref*{#2}\ifthenelse{\equal{#1}{}}{}{(#1)}}}
\newcommand{\fullfigref}[2][]{Figure~\hyperref[#2]{\ref*{#2}\ifthenelse{\equal{#1}{}}{}{(#1)}}}

\title{Supplementary Material: \\ Fine-Structure Qubit Encoded in Metastable Strontium Trapped in an Optical Lattice}

\newcommand{\AffiliationMPQ}{\affiliation{%
  Max-Planck-Institut f{\"u}r Quantenoptik,
  85748 Garching, Germany}}%
\newcommand{\AffiliationMCQST}{\affiliation{%
  Munich Center for Quantum Science and Technology,
  80799 M{\"u}nchen, Germany}}%
\newcommand{\AffiliationLMU}{\affiliation{%
  Fakult{\"a}t f{\"u}r Physik,
  Ludwig-Maximilians-Universit{\"a}t M{\"u}nchen,
  80799 M{\"u}nchen, Germany}}%

\author{S. Pucher}
\author{V. Kl{\"u}sener}
\author{F. Spriestersbach}
\author{J. Geiger}
\AffiliationMPQ{}
\AffiliationMCQST{}
\author{A. Schindewolf}
\author{I. Bloch}
\author{S. Blatt}
\email{sebastian.blatt@mpq.mpg.de}
\AffiliationMPQ{}
\AffiliationMCQST{}
\AffiliationLMU{}

\date{\today}

\maketitle

\section{Fidelity of the Landau--Zener sweep}

We investigated the coherent excitation from $\ket{g}$ to $\ket{\uparrow}$ and its requirements for the optical lattice polarization and quantization axis orientation in a previous publication~\cite{klusener24}.
The Landau--Zener sweep over this transition is realized by performing a linear sweep of the transfer-laser frequency symmetrically over the $\ket{\mathrm{g}}$--$\ket{\uparrow}$ resonance.
We use a scan range of $4$\,kHz and a sweep duration of $50$\,ms corresponding to a ramp speed of 80\,Hz/ms.
To determine the fidelity of the sweep, we analyze the number of atoms $N_\mathrm{rem}$ remaining in $\ket{\mathrm{g}}$ after the scan.
Each of these measurements is normalized to a reference measurement of the initial ground state atom number $N_\mathrm{ref}$.
We define the excitation fidelity as $\mathcal{F}=1-N_\mathrm{rem}/N_\mathrm{ref}$.
Repeating this measurement ten times results in an average excitation fidelity of 97.5(6)\%.

\section{Raman laser system}

For the Raman laser system, we use two diode lasers at wavelengths of $707$\,nm and $679$\,nm, which are locked to the same optical reference cavity using the Pound--Drever--Hall locking scheme~\cite{boyd2024basic}. This ensures phase stability between the two lasers. The cavity has a finesse of $323(3)\times 10^3$ and $222(4)\times 10^3$ at $707$\,nm and $679$\,nm, respectively. We can vary the one-photon detuning $\Delta$ on the GHz scale by locking the laser frequencies to specific longitudinal modes of the cavity, which are separated by the free spectral range of about $1.5\,\mathrm{GHz}$. The two laser beams are spatially overlapped in an optical fiber leading to the experiment chamber.

\section{Branching ratios}

The excited state $\ket{\mathrm{s}}$ decays incoherently to $\ket{\uparrow}$ and $\ket{\downarrow}$ with branching ratios of 21.7\% and 11.6\%, respectively~\cite{heinz20,safronovapriv}.
With a probability of 34.0\%, $\ket{\mathrm{s}}$ decays to the $^3\mathrm{P}_1$ states from where the atoms decay back to $\ket{\mathrm{g}}$~\cite{heinz20,safronovapriv}.
The remaining 32.6\% decay to $m_J \neq 0$ states of the $^3\mathrm{P}_2$ manifold.
To suppress spontaneous decay from $\ket{\mathrm{s}}$ we detune both Raman lasers from the atomic transition frequencies by the one-photon detuning of typically $\Delta \approx -2 \pi \times6$\,GHz.

\section{Data analysis}

We use a CCD camera to measure the absorption of the atoms, which we convert into a number of atoms. We then calculate the mean and standard deviation of the atom number in a $3 \times 3$ pixel region, corresponding to a $10.5 \times 10.5$\,\textmu m$^2$ analysis region.

To correct for slow drifts in the number of atoms loaded into the optical lattice, we interleave the measurements of the Rabi oscillations with reference measurements that are performed without the application of Raman pulses, which we use to determine the atom number in $\ket{\uparrow}$ in $\ket{\uparrow}$ at $t=0$.
We use a linear interpolation of these reference measurements for the correction.
The resulting $N_\uparrow (t)/N_\uparrow (0)$ is the state-preparation-and-measurement error corrected atom number in $\ket{\uparrow}$.
We use the same reference measurements to correct the number of atoms in $\ket{\downarrow}$, given by $N_\downarrow (t)/N_\uparrow (0)$.
To avoid any possible bias introduced by this method, we only use the measured atom number in $\ket{\uparrow}$ to determine the results in the main text.

\section{\label{sec:level1} Excitation fraction and detection fidelity of atoms in $\ket{\downarrow}$}

\begin{figure}[h]
\includegraphics[width=1\columnwidth]{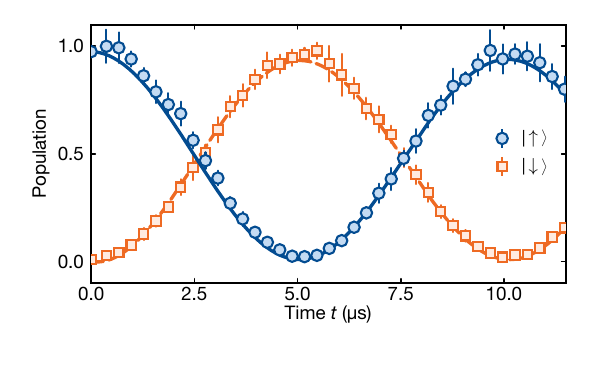}
\caption{Excitation fraction and detection fidelity. We average seven measurements of the Rabi oscillations between  $\ket{\uparrow}$ (blue dots) and $\ket{\downarrow}$ (orange squares). Here, the data is corrected for the Landau-Zener efficiency. We infer the excitation fraction from the number of atoms in $\ket{\uparrow}$ at $t = 5.17$\,\textmu{}s. From the number of atoms in $\ket{\downarrow}$ at this time, we derive the detection fidelity of atoms in $\ket{\downarrow}$.}
\label{fig:rabi_excitation_fraction}
\end{figure}

To determine the maximal $\pi$ pulse excitation fraction, we measure Rabi oscillations with small time steps, as shown in \figref{fig:rabi_excitation_fraction}. Here, we use $\Delta \approx -2 \pi \times6$\,GHz and $\Omega_\uparrow = \Omega_\downarrow \approx 2 \pi \times36$\,MHz. We average seven Rabi oscillations by averaging the number of atoms at every timestep.
Starting in $\ket{\uparrow}$, after a $\pi$ pulse with a time of $t = 5.17$\,\textmu{}s, we find a population of 0.02(1) in $\ket{\uparrow}$, corresponding to an excitation fraction of 98(1)\%.

To estimate the detection efficiency of the atoms in $\ket{\downarrow}$, we use the population in $\ket{\downarrow}$, which we have normalized with the population in $\ket{\uparrow}$. In \figref{fig:rabi_excitation_fraction}, we show this data that is already corrected for the measured Landau--Zener efficiency of 97.5(6)\,\%. At a time $t = 5.17$\,\textmu{}s, we measure a population of 0.94(3) in $\ket{\downarrow}$. From this, we infer a detection fidelity of the atoms in $\ket{\downarrow}$ of $0.94(3)/0.98(1) = 96(3)$\%, where 0.98(1) results from the measured excitation fraction above.

\section{\label{sec:level2} Fit of the Rabi frequency and the envelope of the Rabi oscillations}

\begin{figure}[h]
\includegraphics[width=1\columnwidth]{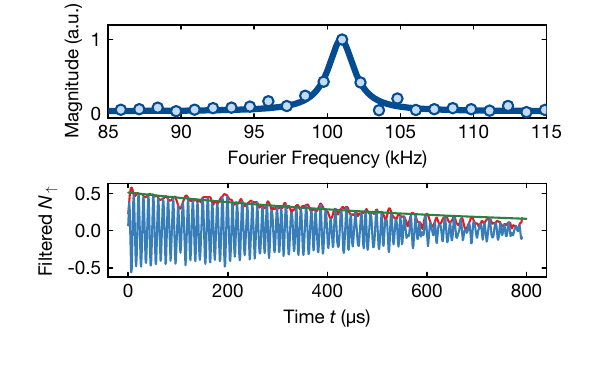}
\caption{Data analysis of the Rabi oscillations. Upper panel: Fourier spectrum of the Rabi oscillations at $\Delta \approx -2 \pi \times$6\,GHz (points). We fit a Lorentzian function to the data to determine the two-photon Rabi frequency $\Omega$ (line). Lower panel: Rabi oscillations after applying a band-pass filter (blue line). We fit the envelope (red line) with an exponential function (green line) to find the decay time of the oscillations.}
\label{fig:data_analysis}
\end{figure}

In our system, we have long-term pulse amplitude fluctuations, which affect the Rabi frequency. As a result, the Rabi frequency drifts and when the shot-to-shot fluctuations get too high, we no longer observe coherent oscillations, but the data points still scatter with the contrast of the oscillations at that time. Thus, we analyze the frequency and the envelope of the oscillations separately.

First, we perform a Fourier transform of the Rabi oscillation data. We fit a Lorentzian function to the resulting Fourier spectrum and use the center of the fit to determine $\Omega$, see upper panel in \figref{fig:data_analysis}. Then, we apply a 2$^{\mathrm{nd}}$-order Butterworth band-pass filter with cut-off frequencies $\pm 50$\% around $\Omega$ to the Rabi oscillations. This suppresses signals with slow and high frequencies and we obtain a sinusoidal function oscillating between -1 and 1, see lower panel in \figref{fig:data_analysis}. To extract the envelope of the oscillations, we use the ObsPy toolbox~\cite{beyreuther2010ObsPy}. This function computes the envelope by first adding the squared amplitudes and its Hilbert transform and then calculating the square root~\cite{kanasewich1981time}. We fit an exponential function to the envelope to extract the decay time $\tau$ of the Rabi oscillations.

In \figref{fig:data_analysis}, we present the results for the Rabi oscillations at $\Delta \approx -2 \pi \times6$\,GHz. We find $\Omega = 2 \pi \times 100.94(4)$\,kHz and $\tau = 684(13)$\,\textmu s. From these results, we infer a $1/e$ damping of the Rabi oscillations of 69(1) cycles.

To demonstrate the agreement of our analysis method with the measured Rabi oscillations, we use the obtained parameters $\Omega$ and $\tau$ in a model, which we compare to the Rabi oscillation data. In addition to the decay of the envelope due to dephasing, this model also takes into account loss out of the $\Lambda$ system due to one-photon scattering.
We model the number of atoms in $\ket{\uparrow}$ as
\begin{equation}
    N_j(t) = 0.5 \cos(\Omega t + \phi_j) e^{-t/\tau} - A ( 1 - e^{-t/\tau_\mathrm{loss}}) + 0.5~,
    \label{eq:damped_oscillations}
  \end{equation}
where $\Omega$ is the carrier frequency, the phase is $\phi_\uparrow=0$ ($\phi_\downarrow=\pi$) for atoms in $\ket{\uparrow}$ ($\ket{\downarrow}$), and we assume an exponential envelope of the oscillations with a decay time of $\tau$. The second term of Eq.~\eqref{eq:damped_oscillations} describes loss out of the $\Lambda$ system due to one-photon scattering with a time scale $\tau_\mathrm{loss}$.
The amplitude $A$ takes into account that atoms that have scattered a photon can be pumped into the dark state of the $\Lambda$ system. Here, we extract $A$ and $\tau_\mathrm{loss}$ from a fit of the difference between the Rabi oscillations and the filtered Rabi oscillations and obtain $\Omega$ and $\tau$ from the analysis described above.
We use these parameters to overlay a curve based on Eqn.~\ref{eq:damped_oscillations} in Fig.~3(d) in the main text, showing the good agreement between the effective and the data.

\section{Measurement of the one-photon scattering rate}

\begin{figure}[h]
\includegraphics[width=1\columnwidth]{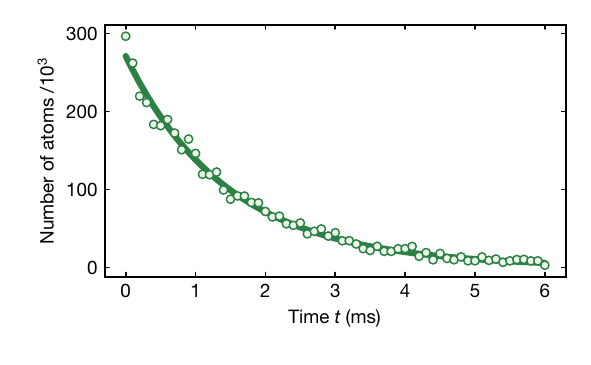}
\caption{Measurement of the scattering rate at $\Delta_\uparrow \approx -2 \pi \times 6$\,GHz and $\Omega_\uparrow \approx 2 \pi \times36$\,MHz. We prepare the atoms in $\ket{\uparrow}$ and turn on the up laser. Then, we measure the number of atoms in $\ket{\uparrow}$ for various excitation times. With a fit of a rate equation model, we find the scattering rate of the up laser.}
\label{fig:scattering_rate}
\end{figure}

One-photon scattering of the Raman lasers leads to decoherence in our system. Here, we discuss our measurements of the one-photon scattering rates. To characterize the scattering on the $\ket{\uparrow}$--$\ket{s}$ transition, we prepare the atoms in $\ket{\uparrow}$. Then, we turn on just the up laser. After various excitation times, we measure the number of atoms in the initial state. When the atoms scatter a photon, they can decay to $^3\mathrm{P}_0$, $^3\mathrm{P}_1$, and to the five Zeeman substates of the $^3\mathrm{P}_2$ manifold.
To include these effects and get an estimate of the scattering rate, we fit a rate equation model to our data. This model includes all Zeeman substates of the relevant states. In \figref{fig:scattering_rate}, we present a measurement of the scattering rate at $\Delta_\uparrow \approx -2 \pi \times 6$\,GHz with $\Omega_\uparrow \approx 2 \pi \times36$\,MHz. From the fit, we extract a scattering rate of $867(19)\,\mathrm{s}^{-1}$ and we infer a maximal decay time of the Rabi oscillations of $1.15(2)$\,ms. We assume that $|\Delta_\uparrow|$ is much larger than $\Omega_\uparrow$ and the natural linewidth, and fit a curve $a\Delta_\uparrow^2$ to the resulting scattering-limited decay times shown in Fig.~3(e) of the main text, and we find a coefficient $a = 39(2)$\,\textmu{}s$/(2\pi\times\mathrm{GHz})^2$.

\section{Atomic polarizabilities}

\begin{figure}
\includegraphics[width=1\columnwidth]{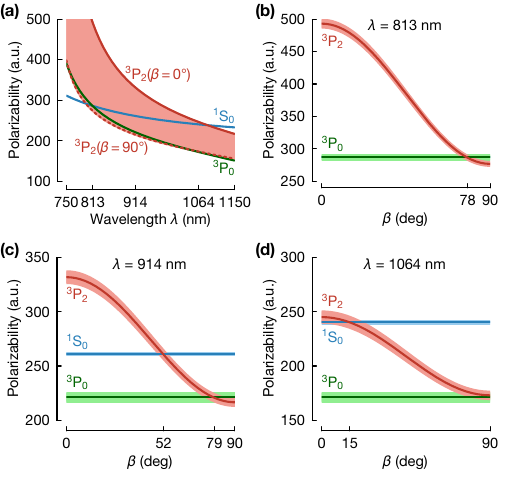}
\caption{Atomic polarizabilities. (a) Theoretically calculated polarizability of the $^1\mathrm{S}_0$ (blue), $^3\mathrm{P}_0$ (green) and $^3\mathrm{P}_2$ (red) states as a function of trap wavelength. For the $^3\mathrm{P}_2$ state the polarizability can be tuned within the shaded region bounded by the limiting cases of parallel (solid line) and orthogonal (dashed line) trap polarization relative to the quantization axis. (b),(c),(d) Polarizability as a function of the angle $\beta$ between trap polarization and quantization axis for wavelengths of 813\,nm, 914\,nm, and 1064\,nm, respectively. The shaded regions indicated one-sigma confidence intervals on the theoretical polarizability values, which give rise to a substantial uncertainty for the predicted magic angles.}
\label{fig:polarizability}
\end{figure}

\fullfigref[a]{fig:polarizability} shows the polarizability of the qubit states $^3\mathrm{P}_0$ and $^3\mathrm{P}_2$ $(m_J = 0)$ used in this work, as well as the polarizability of the ground state $^1\mathrm{S}_0$ as a function of trap wavelength $\lambda$~\cite{trautmann23}. The polarizability of the states $^1\mathrm{S}_0$ and $^3\mathrm{P}_0$ is solely determined by $\lambda$ due to the lack of angular momentum ($J=0$). In contrast, the polarizability of the nonspherical $^3\mathrm{P}_2$ state ($J=2$) has an additional dependence on the trap polarization relative to the quantization axis. For the $^3\mathrm{P}_2$ ($m_J=0$) state used in this work, the total polarizability is given by~\cite{lekien13}
\begin{equation}
\label{eq:polarizability}
  \begin{aligned}
    \alpha(\lambda, \beta)& =  \alpha_{\mathrm{s}}(\lambda) - \alpha_{\mathrm{t}}(\lambda) \dfrac{3 \cos^{2}\beta -1 }{2}~,\\
      \end{aligned}
\end{equation}
where $\alpha_{\mathrm{s}}$ and $\alpha_{\mathrm{t}}$ denote scalar and tensor polarizabilities, respectively, and $\cos\beta$ is the projection of the polarization vector onto the quantization axis. The shaded region in \figref[a]{fig:polarizability} indicates the tuning range of the $^3\mathrm{P}_2$ polarizability bounded by the limiting cases of trap polarization parallel ($\beta=0\degree$) or orthogonal ($\beta=90\degree$) to the quantization axis. For $\beta=90\degree$, the differential polarizability between the $^3\mathrm{P}_0$ and $^3\mathrm{P}_2$ states is very small over a large range of wavelengths. In this regime the existence of a magic trapping condition with (near-)perfect differential light shift cancellation is possible. Because the uncertainty on the theoretically calculated polarizabilites is on the order of 1-2\% and the crossings are very shallow, the uncertainty on predicted magic wavelengths and angles is substantial and experimental input is required.

Figures~\hyperref[fig:polarizability]{\ref*{fig:polarizability}(b)}--\hyperref[fig:polarizability]{\ref*{fig:polarizability}(d)} show polarizability data as a function of the angle $\beta$ including theoretical confidence intervals for three specific wavelengths of interest~\cite{safronovapriv}. Trap wavelengths of 914\,nm and 1064\,nm are used in this work for the horizontal and vertical optical lattices, respectively. The magic wavelength of 813\,nm is of particular interest because here the differential light shift on the $^1\mathrm{S}_0$--$^3\mathrm{P}_0$ clock transition is canceled, and obtaining a triple-magic wavelength in combination with the $^3\mathrm{P}_2$ state might be feasible. Experimentally, we determine magic polarization angles by maximizing the Ramsey contrast by scanning the lattice polarization angles. At 914\,nm, we observe a local maximum in the Ramsey contrast for an angle $\beta$ slightly above the predicted magic angle of $\beta_\mathrm{m,914}=79\degree$. A precise determination of the magic angle is beyond the scope of this work, but would be a fruitful topic for future experiments. At 1064\,nm, the differential light shift is minimized for $\beta=90\degree$, but no magic trapping conditions could be achieved. These results indicate that an upper bound on the trap wavelength for realizing a magic trap for the fine-structure qubit is located between 914\,nm and 1064\,nm. Observing the scaling of the polarizability of the
$^3\mathrm{P}_0$ and $^3\mathrm{P}_2$ states with wavelength in \figref[a]{fig:polarizability} suggests that the existence of a magic angle at 914\,nm supports the existence of a triple-magic trapping condition for the $^1\mathrm{S}_0$--$^3\mathrm{P}_0$--$^3\mathrm{P}_2$ qutrit at 813\,nm.

% \bibliography{finestructure}